\documentclass[biblatex, english]{lni}
\usepackage{graphicx}
\usepackage{subcaption}
\usepackage{multirow}
\usepackage{tikz}
\usepackage[T1]{fontenc}
\usepackage{longtable}
\usepackage{pgf}
\usepackage{color, colortbl}
\usepackage{import}

\definecolor{orange}{rgb}{
    1, 0.654, 0.169
} 

\definecolor{blue}{rgb}{
    0.282, 0.624, 0.710
} 

\newcommand{\stockrow}{
    \rowcolor{blue}
}

\newcommand{\stockcell}{
    \cellcolor{blue}
}

\newcommand{\olrow}{
    \rowcolor{orange}
}

\newcommand{\olcell}{
    \cellcolor{orange}
}

\AtEndPreamble{%
  \usepackage{cleveref}
}

\DeclareMathOperator{\Orderline}{Orderline}

\DeclareMathOperator{\Stock}{Stock}

\DeclareMathOperator{\JOIN}{\bowtie}

\begin{document}

\title[Indexing Join Inputs]{Indexing Join Inputs for Fast Queries and Maintenance}
\author[1]{Wenhui Lyu}{wenhui@cs.wisc.edu}{0009-0008-0946-4414}
\author[2]{Goetz Graefe}{goetzg@google.com}{0000-0003-0194-6466}
\affil[1]{University~of~Wisconsin--Madison, USA}
\affil[2]{Google, Madison~Wis., USA}
\maketitle

\begin{abstract} 
    In database systems, joins are often expensive despite many years of research producing numerous join algorithms.
    Precomputed and materialized join views deliver the best query performance, whereas traditional indexes, used as pre-sorted inputs for merge joins, permit very efficient maintenance.
    Neither traditional indexes nor materialized join views require blocking phases, in contrast to query-time sorting and transient indexes, e.g., hash tables in hash joins, that impose high memory requirements and possibly spill to temporary storage.

    Here, we introduce a hybrid of traditional indexing and materialized join views.
    The \textit{merged index} can be implemented with traditional b-trees, permits high-bandwidth maintenance using log-structured merge-forests, supports all join types (inner joins, all outer joins, all semi joins), and enables non-blocking query processing.
    Experiments across a wide range of scenarios confirm its query performance comparable to materialized join views and maintenance efficiency comparable to traditional indexes.
\end{abstract}


\section{Introduction}\label{sec:intro}

In database systems, computing joins is an important and well-studied topic.
Hash joins~\cite{dewitt1985multiprocessor,dewitt1984main,kim1980product} and sort-merge joins~\cite{blasgen1977storage} have been considered the algorithms of choice, but they require blocking phases to build hash tables or sort inputs.
These blocking phases are memory-intensive and may spill to temporary external storage, causing performance degradation.
Index nested-loops joins do not require blocking phases but are often much less efficient than hash joins and sort-merge joins~\cite{Haas1999RippleJoins}.
Proposed remedies for hash joins include symmetric hash joins~\cite{hong1993shj, wilschut1991shj, Urhan2000XJoinAR, lawrence2005early}.
Here, we propose revisiting index-based merge joins.
Join inputs indexed on their join columns are effectively pre-sorted inputs for merge joins and do not require blocking phases before producing the first tuple.
Index-based merge joins, with performance improvement proposed here, become viable for non-blocking join processing.

Precomputing and storing join outputs as materialized views allows join queries to be answered more efficiently than indexing join inputs, also without blocking phases.
However, in databases with constant updates, materialized join views require efficient maintenance algorithms.
Despite extensive research efforts and considerable progress from classic techniques for incremental view maintenance (IVM)~\cite{Idris2017DyanmicYanakakis, GUPTAchangetable, larson2007outerjoin, dbtoasterFK, Blakeley1990PerformanceAnalysis, wilschut1991shj, Urhan2000XJoinAR, hong1993shj}, commercial database systems typically provide limited support for efficient incremental maintenance of join views, especially of outer joins~\cite{oracle2024mvrefresh, postgresql2024incremental, ibm2024mqts,mysql2024viewupdatability, microsoft2024indexedviews, databricks2018stream}.
Many commercial database systems provide no more than indexes on join input tables including efficient maintenance based on extensive research and development~\cite{ibm2024indexsync, microsoft2024modifyindex, oracle2024index}.

An alternative to traditional materialized views avoids materializing the join result, instead storing an alternative data structure or ``proxy store'' balancing efficient generation of the join result with efficient incremental maintenance~\cite{Blakeley1990PerformanceAnalysis, oneil1995joinindex, Idris2017DyanmicYanakakis}.
An ideal proxy store enables generation of query results as efficiently ``as if they were materialized''~\cite{Idris2017DyanmicYanakakis}, maintenance as efficient as traditional single-table indexes, and query coverage including all outer joins and semi joins.
In this sense, traditional materialized views are ideal in only the first one of these three dimensions,
whereas a pair of single-table indexes is ideal only in the other two dimensions.

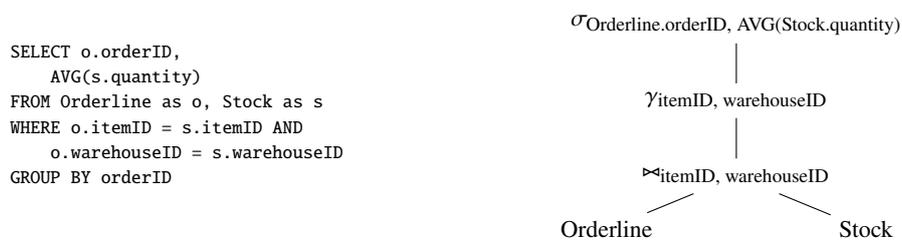
\begin{figure}[htb]
    \begin{minipage}{0.6\linewidth}
        \footnotesize
        \begin{verbatim}
SELECT o.orderID,
    AVG(s.quantity)
FROM Orderline as o, Stock as s
WHERE o.itemID = s.itemID AND
    o.warehouseID = s.warehouseID
GROUP BY orderID
        \end{verbatim}
    \end{minipage}
    \begin{minipage}{0.38\linewidth}
        \begin{tikzpicture}[node distance=1cm]
            \node (proj) {\(\sigma_\text{Orderline.orderID, AVG(Stock.quantity)}\)};
            \node (groupby) [below of=proj] {\(\gamma_\text{itemID, warehouseID}\)};
            \node (join) [below of=groupby] {\(\JOIN_\text{itemID, warehouseID}\)};
            \node (scan1) [below left of=join, xshift=-1cm] {Orderline};
            \node (scan2) [below right of=join, xshift=1cm] {Stock};

            \draw[-] (proj) -- (groupby);
            \draw[-] (groupby) -- (join);
            \draw[-] (join) -- (scan1);
            \draw[-] (join) -- (scan2);
        \end{tikzpicture}
    \end{minipage}
    \caption{Example: A join operator with additional query processing steps.}\label{fig:join-operator}
\end{figure}

In contrast to the two traditional alternatives,
this paper introduces the merged index~\cite{graefe2007merged} as an ideal proxy store for materialized join views in all three dimensions: query performance comparable to materialized join views, maintenance efficiency comparable to traditional indexes, and a foundation for inner, outer, and semi joins.
By providing an efficient way of indexing inputs of one join operator, a merged index is a proxy store of a simple binary join only.\footnote{
    But such an optimized join operator can be an intermediate step in a larger query such as \cref{fig:join-operator}.
    We leave proxy stores that capture complex views for future work.
}
In addition, the merged index can be implemented with traditional b-trees, which are ubiquitous in commercial database systems, or with the widely supported log-structured merge-forests, which offers high-bandwidth update performance; our experiments evaluate both storage formats.
The merged index adds little space overhead to equivalent indexes and is a surprisingly natural compression of a materialized join view.

Our contributions include the following:
\begin{itemize}
    \item To database query processing, we contribute by optimizing index-based merge joins and providing an efficient non-blocking join algorithm.
    \item To indexing techniques, we contribute by
    \begin{itemize}[noitemsep,topsep=0pt,parsep=0pt]
        \item experimentally comparing traditional indexes, traditional materialized views, and merged indexes;
        \item integrating merged indexes with log-structured merge-forests;
        and
        \item introducing a simple solution that combines ideal query performance, ideal update performance, and coverage of all join types.
    \end{itemize}
    \item To materialized views, we contribute by optimizing index-based merge joins such that the merged index becomes an ideal proxy store for and an alternative to traditional storage for materialized join views.
\end{itemize}

In the following sections, \cref{sec:background} provides background on index-based join plans, materialized views and incremental maintenance techniques, and storage formats; whereupon \cref{sec:prior} reviews recent studies on non-blocking joins and proxy stores for materialized join views.
\Cref{sec:techniques} details the design and trade-offs of merged indexes and their application to binary joins.
In \cref{sec:exp}, experiments compare the performance of merged indexes, indexed base tables, and materialized joins, supporting our central claim that the merged index matches traditional indexes on maintenance efficiency and materialized join views on query efficiency.
\Cref{sec:sum} summarizes our findings and offers conclusions.

\section{Background and Preliminaries}\label{sec:background}

This paper focuses on optimizing a join of two tables, with the join output potentially serving as intermediate steps in larger queries.
For instance, consider a query computing the average stock quantity of items in the same order as shown in \cref{fig:join-operator}.
Additional processing steps, grouping and projection, recompute fully for each query, while the merged index accelerates the join step.
Similarly, query engines process preceding steps and prepare input tables for the join.
Another join may exist in the preceding or subsequent steps; query engines may choose an appropriate plan for it which may be a merged-index-backed join or not.
Proxy stores that cache more complex views such as multi-way joins are left for future work.

In this context, \cref{ssec:index-join} reviews index-based merge joins and their benefits.
\Cref{ssec:classic-ivm} discusses the consistency between materialized views and base tables and how state-of-the-art IVM techniques for joins are often unable to keep join views in sync at all times.
Index maintenance, on the other hand, offers strict consistency with base tables in most commercial systems and, therefore, sets the standard for ideal maintenance.
As one starting point of this study is consideration of memory hierarchies and support of external storage, \cref{ssec:storage} covers storage formats for indexes on external devices.

\subsection{Index-based merge joins}\label{ssec:index-join}

Earlier studies on non-blocking join algorithms include symmetric hash joins~\cite{hong1993shj,wilschut1991shj, Urhan2000XJoinAR, lawrence2005early}, which keep hash tables on both join inputs for reuse and implement a non-blocking version of hash joins.
In addition, ripple joins~\cite{Haas1999RippleJoins, Luo2002hashripple} are based on index nested-loops joins.
Here, we focus on merge joins using indexes on join inputs.

Traditional sort-merge joins require blocking phases to sort their inputs.
Indexed join inputs (typically in the format of b-trees) retain and cache the efforts of this sort phase in a durable and maintainable format.
These pre-sorted inputs can directly participate in a merge join.

Covering indexes\footnote{Covering indexes include all primary indexes (also known as clustered indexes and index-organized tables) and, depending on a specific query, can include secondary indexes (also known as non-clustered indexes or images).} ``cover'' all required columns by a join query, enabling index-only retrieval.
Non-covering indexes require back-joins to the base tables for the missing columns.
Even with covering indexes, such a merge join may not answer join queries as efficiently as directly querying a precomputed join view.

In addition to avoiding blocking phases, index-based merge joins offer the benefits of
\begin{itemize}[noitemsep,topsep=0pt,parsep=0pt,partopsep=0pt]
    \item rendering sorted join output, which is potentially useful for a subsequent sort-based operator; and
    \item supporting inner joins, all outer joins, and all semi joins of the same input tables on the same columns\footnote{Or, in the case of composite keys, any prefix.} because all rows including non-matching ones are stored in these indexes.
\end{itemize}

Additionally, indexes and their sort order offer the following benefits:
\begin{itemize}
    \item Compatibility with a memory hierarchy including external storage.
    \item Robust and evolving ecosystems: B-trees, the typical storage format of indexes, are ubiquitous in database systems and with well-known techniques for concurrency control in transactions and efficient maintenance.
    Ongoing research and development have been devoted to relevant topics, such as efficient index maintenance and other storage formats, including log-structured merge-forests~\cite{rocksdb21,wisckey,leanstore18,lepers2019kvell,dayan2017monkey}.
    \item Potential participation in other queries and index operations such as set operations (e.g., intersection and union).
\end{itemize}

\noindent These benefits are inherited by the merged index and add to the strengths of our approach.

\subsection{Materialized views and the problem of inconsistency}\label{ssec:classic-ivm}

Data warehouses often permit materialized views to lag behind their base tables instead of enforcing immediate synchronization within each update transaction.
When full recomputation is the only method for view maintenance, materialized views remain stale except immediately after a scheduled maintenance.\footnote{
    For instance, \citet{ibm2024mqts,postgresql2024incremental} support only full recomputation for materialized join views.
    So does \citet{mysql2024viewupdatability} unless the join is a primary-key foreign-key inner join and only the table with the primary key is updated.
}
However, there are many scenarios in which only up-to-date materialized views are useful.
For example, if a materialized view accelerates predicate evaluation for a query but lacks some columns needed in the query, back-joins to the base table are required---though likely incorrect if the materialized view and base table are inconsistent.
Moreover, if query optimization chooses automatically whether a query plan uses the materialized view, seemingly equivalent plans produce different results, confusing users.
One solution to this inconsistency offers the simplest semantics and is most predictable for users: keeping all base tables, their indexes, and dependent materialized views up-to-date at all times.

Incremental view maintenance (IVM) avoids expensive full recomputation by incrementally applying database updates to materialized views~\cite{gupta1995maintenance,Zhuge1995ViewMaintenance,Blakeley1990PerformanceAnalysis}.
Let \(R \JOIN S\) denote a materialized view of a binary equality join between \(R\) and \(S\), and \(\Delta R\) and \(\Delta S\) the updates to \(R\) and \(S\).\footnote{
    \(\Delta R\) consists of insertions, updates, and deletes. One possible way to distinguish the three update types is to use a positive count for insertions and a negative count for deletions, and updates are expressed as deletions followed by insertions~\cite{mumick1997maintenance,GUPTAchangetable,mv12,dbtoasterFK}.
}
To incrementally maintain \(R \JOIN S\) with changes \(\Delta R\) and \(\Delta S\) requires three joins, two of which involving full relations: \(\Delta R \JOIN S\), \(R \JOIN \Delta S\), and \(\Delta R \JOIN \Delta S\).
\citet{Budiu2023DBSP} describe this as the ``bi-linear'' nature of joins.
For outer joins, prior studies have introduced incremental maintenance techniques with performance comparable to or slightly worse than inner joins~\cite{GUPTAchangetable, larson2007outerjoin}, which also exhibit a bi-linear cost.
However, most commercial systems~\cite{mysql2024viewupdatability, microsoft2024indexedviews, oracle2024mvrefresh, databricks2018stream} and research prototypes~\cite{dbtoaster} provide limited or no IVM support for materialized outer joins.
Higher-order IVM of DBToaster~\cite{dbtoaster} greatly optimizes classic IVM techniques in propagating changes through complex views such as multi-way joins at the cost of large space overhead (mitigated by F-IVM~\cite{f-ivm}).
However, these techniques are less meaningful in optimizing simple binary joins (there are as high as two orders in total) and remain constrained by the bi-linear cost of joins except in certain cases\footnote{
    \Citet{dbtoasterFK} introduced an optimization for the following case and avoided evaluating two full relations: joins with foreign-key constraints, with the requirement that the relation providing the primary key only undergo insertions and deletions without updates~\cite{dbtoasterFK}.
    Example maintenance in our experiments includes a case of primary-key foreign-key joins but violates the update restriction.
}.

The classic IVM design consists of three stages: (a) collecting the delta of base tables (``base delta''), (b) deriving the delta of the view (``derived delta''), and (c) refreshing the view with its derived delta.
\Citet{GUPTAchangetable, dbtoasterFK} create a ``summary delta'' or ``change table'' to batch derived delta before merging it with the view, deferring stage (c).
Materialized view logs by \citet{oracle2024mvrefresh} batch base deltas before entering stage (b).
Difference traces from \citet{mcsherry2013differentialdataflow} defer both stages (b) and (c).
This deferral semantics introduces performance optimization by merging changes and batching computation; \citet{mcsherry2013differentialdataflow} also use difference traces to provide versions over multiple time windows.
However, this approach may not suit workloads requiring strict transactional consistency, such as queries that back-join a materialized view with base tables.\footnote{
    Opting out of deferral is possible and implemented by \citet{oracle2024mvrefresh} at a higher cost and without support for outer joins.
}

In contrast, keeping secondary indexes up-to-date at all times with base tables is widely supported in all transactional database systems~\cite{oracle2024index, microsoft2024modifyindex}, including systems lacking IVM support~\cite{postgres2024index, ibm2024indexsync}.
Unlike the bi-linear cost of IVM for joins, only \(\Delta R\) must be reflected in a traditional index on join input \(R\).
Index maintenance remains efficient even as part of a transaction, ensuring strict consistency when needed.
It sets the standard for ideal maintenance.
Existing infrastructure with this ideal maintenance can also support merged indexes, offering efficient maintenance and options for strict consistency with base tables.

\subsection{External storage formats for indexes}\label{ssec:storage}

Scalability to very large databases, memory hierarchies, and comprehensive support for external storage are an important consideration in our work.
In contrast, many related studies~\cite{Idris2017DyanmicYanakakis, Berkholz2017conjunctive, mcsherry2013differentialdataflow, dbtoaster} focus on or prioritize main-memory algorithms.

The classic and ubiquitous index format in databases is the b-tree~\cite{bayer1972organization, comer1979ubiquitous, Graefe11a, Graefe24}.
The merged index in b-tree format illustrates that the concepts of an index and of a b-tree structure are not the same:
A single b-tree or a single log-structured merge-forest can map keys to values for multiple tables, i.e., store multiple indexes.
Efficient maintenance techniques for b-trees include sorting index entries before insertion, similar to efficient index creation by first sorting future index entries, though the latter is much more widely known and used than the former.

For even more efficient maintenance, log-structured merge-forests~\cite{ONeil1996LSMTree} and stepped-merge forests~\cite{Jagadish1997LSM} run external merge sort ``on a funny schedule'':\footnote{
    Martin Kersten once agreed that database cracking~\cite{tozun2024reminiscences,DatabaseCracking} is ``quicksort on a funny schedule''.
}
Changes are initially absorbed in an in-memory b-tree, which spills into external memory-sized initial partitions.
Subsequent ``compaction'' steps merge these partitions into larger b-trees.
In other words, log-structured merge-forests delay tree optimization as part of a continuous index creation process.
Notably, they do not delay index maintenance or the capture of new index contents but only the optimization of the index structure.
This approach complements the techniques introduced here by enabling efficient capture of incremental index changes.
However, log-structured merge-forests may slow down queries due to the need to scan or search multiple partitions.

\section{Related Work}\label{sec:prior}

This section reviews non-blocking join algorithms and proxy stores for materialized join views proposed in recent studies.

\paragraph{Non-blocking join algorithms:}

Recent work on non-blocking join algorithms~\cite{Dittrich2002PMJ, Chen2010prjoin, Urhan2000XJoinAR, mokbel2004hash} focuses on streaming workloads and balances three aspects: (a) response time of first result rows, (b) continuous query processing despite data source delays, e.g., over a network, and (c) end-to-end execution time.
These objectives and contexts differ from ours.
Our work does not address aspect (b) but introduces additional objectives: using the intermediate storage structure of non-blocking joins, the merged index, as an optimal proxy store for the corresponding materialized join view.

\paragraph{Proxy stores for materialized join views:}

An alternative to IVM is maintaining a proxy store instead of a full materialized view.
Proxy stores balance maintenance efficiency with the efficiency to generate the materialized view or a queried subset of it.
Join indexes~\cite{oneil1995joinindex,valduriez87joinindex}, symmetric hash tables~\cite{wilschut1991shj, hong1993shj, Urhan2000XJoinAR, lawrence2005early}, and traditional indexes on join inputs are examples of proxy stores for materialized join views.
\Citet{Idris2017DyanmicYanakakis, Berkholz2017conjunctive} propose proxy stores for a generic class of queries, which are also applicable to binary joins.
They provide theoretical guarantees under assumptions of main-memory data structures, which represent a different approach from ours.

In a join index, each record consists of the join key and row IDs from input tables.
Join indexes are similar to merged indexes in precomputing and storing the matching status of join inputs in one storage structure.
In contrast, merged indexes are more versatile:
Like materialized join views, merged indexes allow additional columns to be included.
These additional columns can cover a join query and enable index-only retrieval, achieving query performance comparable to materialized join views.
Join indexes, however, must back-join with base tables to retrieve columns selected by join queries.

Experiments in \cref{sec:exp} focus on the three dimensions of ideal proxy stores as discussed in \cref{sec:intro}.
These dimensions are also valid reference points for all proxy stores and likewise all non-blocking join algorithms.

\section{Techniques}\label{sec:techniques}

\subsection{The original design of the merged index}\label{ssec:merged-index-design}

The merged index~\cite{graefe2007merged} is a traditional storage structure, e.g., a b-tree or a log-structured merge-forest~\cite{Jagadish1997LSM,ONeil1996LSMTree}, unusual only by storing and interleaving multiple record types.
One common use case is for a pair of database tables with shared columns, typically the primary key of one table and the foreign key of the other.
In our design, a merged index stores two single-table indexes of those tables.
It interleaves their records in one sort order using the shared columns as the leading sort columns (also known as ``master-detail clustering''~\cite{graefe2007merged}).

As an example based on the TPC-C benchmark~\cite{tpc2010benchmark}, a merged index organized by (warehouseID, itemID) interleaves index entries from the primary index of the stock table and the secondary index of the orderline table.
\Cref{tab:merged-index} shows how a merged index stores these two tables: the stock table has two rows, each with a unique primary key, and the orderline table has four rows.
They add up to 6 index entries in the merged index.

\begin{table}
    \caption{A merged index that interleaves the stock records (two blue rows) and orderline records (four orange rows) by (warehouseID, itemID).}\label{tab:merged-index}
    \centering
    \begin{tabular}{c c | c | c}
        \hline
        \textbf{warehouseID} & \textbf{itemID} & \textbf{Source Index} & \textbf{Payload} \\
        \hline
        \stockrow
        1 & 1 & Stock & \(s_1\) \\
        \olrow
        1 & 1 & Orderline & \(ol_1\) \\
        \olrow
        1 & 1 & Orderline & \(ol_2\) \\
        \stockrow
        1 & 2 & Stock & \(s_2\) \\
        \olrow
        1 & 2 & Orderline & \(ol_3\) \\
        \olrow
        1 & 2 & Orderline & \(ol_4\) \\
        \hline
    \end{tabular}
\end{table}

The example record structure displayed in \cref{tab:merged-index} leads with the shared sort columns: (warehouseID, itemID).\footnote{
    Different from the original design, each sort column does not have a prefix ``domain tag'' here.
    Such a domain tag allows multiple schemas for the leading key to be stored in the same merged index.
    For example, the leading key of an order record can be (warehouseID, itemID, orderID), and the leading key of a customer record can be (warehouseID, itemID, customerID).
    Different domain tags of the third column naturally separate customer record and order records as two groups.
    It also allows each record to be looked up by their key without ambiguity even if an orderID collides with a customerID.
    For the scope of this paper, the design that two tables share a single schema for the leading key is sufficient.
}
The next column is the ``source index'' column, providing an identifier for the source single-table index.
For a merged index with only two tables, this field can be as simple as a boolean value that takes up one bit.
Techniques such as folding this bit into adjacent columns can avoid such an overhead entirely.\footnote{
    Our implementation on existing key-value stores~\cite{leanstore18,rocksdb21} avoids the one-bit overhead.
    Leanstore stores the key length and the value length of each key-value pair even for single-table indexes, which suffices as the index identifier for our use.
    If two tables happen to have the same key and value lengths, the index identifier can also be folded into the these length fields, as they typically do not take up all allocated bits.
}
Such a merged index should introduce virtually zero space overhead compared to the sum of equivalent indexes on the base tables.

The merged index may more common in pre-relational database management, e.g., hierarchical~\cite{Domdouzis2021} and network databases~\cite{Lake2013}, than in relational databases.
\Citet{haerder1978access} explored a similar idea under the name ``combined image'' in the context of relational database systems.
Clustering in object-oriented database systems~\cite{DARMONT199655} is also related.
Among post-relational and no-SQL systems, Cassandra's wide partitions~\cite[91]{carpenter2022cassandra} is another related pattern.
A wide partition stores data with the same partition key together, although they potentially have different schemas.
Flume's \texttt{join()}, another related idea, returns the ``co-group-by''~\cite{beam2024cogroupbykey} result instead of a flat join table; the co-group-by result follows a schema of (join key, (values from table 1, values from table 2)).

\subsection{Trade-offs of merged indexes}

\Citet{graefe2007merged} explains relevant details and trade-offs of the merged index.
Concurrency control and recovery for merged indexes are similar to those for traditional indexes with minor complications\footnote{
     For example, the key-range locking logic may exhibit unexpected behavior for a merged index: locking an orderline record may also lock the next stock record which has a different sort key.
    Existing techniques have provided solutions to this problem~\cite{graefe2007locking}.
}.
Major trade-offs of merged indexes include the following:

\paragraph{Higher cost of dropping or adding an entire index.} Unlike traditional indexes that can be dropped or added independently, a merged index must perform bulk deletion or insertion operations. Much like ``partitioned b-trees'' in the original design~\cite{graefe2007merged}, log-structured merge-forests~\cite{ONeil1996LSMTree,Jagadish1997LSM} can mitigate this cost: Sorting inserted entries or consolidating tombstone records is deferred until and ``hidden'' by the next compaction run. 
Before such a compaction, the loaded index resides in a transiently independent ``partition'' but is immediately available for queries.
Our experiments simulate this scenario by loading two indexes into a log-structured merge-forest without sorting them first, with results presented in \cref{fig:lsm_load_time}.
 
\paragraph{While a merged index can provide records from only one table, it is not as efficient as a single-table index for this purpose.}
If performance of such a query is critical, an additional single-table index can be created instead of a merged index. 
Such a single-table index can also exist alongside the merged index if the workloads justify the use of both. 

\subsection{Incremental maintenance of merged indexes}

Maintenance of a merged index is very similar to maintenance of the traditional b-trees that are single-table indexes.
In the case of an insertion to any base table stored in the merged index, the corresponding entry is inserted into the merged index.
Similarly, deletions and updates are reflected in the merged index in a one-to-one manner.
Compared with the bi-linear cost of materialized join views (\cref{ssec:classic-ivm}), an update does not need to be evaluated against the other table and can be directly applied to the merged index.

A merged index can be easily supported on existing database infrastructure, such as b-trees or log-structured merge-forests.
It also inherits existing mechanisms for concurrency control, recovery, and index de-fragmentation.
Transaction costs (locking, logging, etc.) in merged indexes are also similar as those in traditional indexes, with minor complications as discussed above.
Drawbacks compared to single-table indexes (the higher cost of dropping or adding an entire new index and extracting records from only one table) do not affect efficiency of incrementally maintaining the merged index.
Our experiments in \cref{sec:exp} further support that merged indexes are as efficient as traditional indexes in maintenance against point read-write transactions that update the base tables.

\subsection{Merged index for binary joins}

\subsubsection{Physical de-normalization and natural compression of join results}

Merged indexes ``shift de-normalization from the logical level of tables and rows to the physical level of indexes and records''~\cite{graefe2007merged}.
In the example of \cref{tab:merged-index}, the merged index effectively precomputes the join result of the stock and orderline tables.
However, the merged index does not denormalize matching records as the equivalent join view (\cref{tab:join}) does.
The join view has 4 longer rows, as each stock row is duplicated as many times as it is matched with orderline rows.
Instead, the merged index simply interleaves matching records still in their normalized forms.
Such a physical storage enables efficient retrieval of join results and remains akin to a precomputed join.
Because of physical rather than logical denormalization, the merged index stores one copy of each record even if it is matched multiple times.
It is thus a natural compression of the join result.

\begin{table}[!htb]
    \caption{The join result of the stock and orderline tables.}\label{tab:join}
    \centering
    \begin{tabular}{c c | c | c}
        \hline
        \textbf{warehouseID} & \textbf{itemID} & \textbf{Stock} & \textbf{Orderline} \\
        \hline
        1 & 1 & \stockcell \(s_1\) & \olcell \(ol_1\) \\
        1 & 1 & \stockcell \(s_1\) & \olcell \(ol_2\) \\
        1 & 2 & \stockcell \(s_2\) & \olcell \(ol_3\) \\
        1 & 2 & \stockcell \(s_2\) & \olcell \(ol_4\) \\
        \hline
    \end{tabular}
\end{table}

Factorized databases~\cite{factorizedDB} also avoid denormalizing join matches into a join result but replace the flat table representation with a more compact factorized representation.
Such a representation is primarily on the logical level; their provided implementation~\cite{f-ivm} is in-memory and stores each factorized view in a separate multi-index map.
In contrast, a merged index, durable on external storage, interleaves two record types in a sort order and optimizes disk accesses for records with matching join keys.
We encourage readers to explore the viability of using merged indexes to implement factorized databases.

\subsubsection{Speeding up join queries}

A merged index consisting of records from two tables and organized by join columns can be used to answer binary join queries efficiently.
Let's consider the following simple binary join query:

\begin{verbatim}
SELECT *
FROM Orderline as o, Stock as s
WHERE o.itemID = s.itemID AND
    o.warehouseID = s.warehouseID
    AND warehouseID = 1 AND itemID = 2
\end{verbatim}

If the full view \(\Stock \JOIN \Orderline\) is materialized (\cref{tab:join}), the query can be answered by a point lookup in such a view.
This point lookup returns the last two rows, \(\left(s_2, ol_3\right)\) and \(\left(s_2, ol_4\right)\).
Directly querying this view sets the standard for ideal query performance.

With two traditional indexes, answering this query requires one point lookup in each index.
Stock rows with required warehouseID and itemID are found to be \(s_2\), and orderline rows are found to be \(ol_3\) and \(ol_4\).
Then, a Cartesian product is computed over found matches.

In contrast to traditional indexes, the merged index stores all matching records together, enabling a single index search to retrieve the result.
For such a lookup, the increase in tree depth due to storing both tables in one b-tree is logarithmic and renders the additional key comparisons very moderate if not negligible.
Merged indexes also require computing the Cartesian product, adding a moderate computation overhead but no additional IO cost compared to a materialized join.
In fact, the compression effect of the merged index may result in a smaller IO cost than a materialized join view because of fewer comparisons needed to traverse the index plus reduced scan volume.
Our experiments in \cref{sec:exp} further support that merged indexes are as efficient as materialized join views in queries.

\subsubsection{How a query planner selects a merged-index-backed join}

Storing multiple record types in one storage structure is a tested technique in commercial databases, including \citet{oracle-table-cluster,Google-Spanner}, and can work well with a query planner.
After all, as an innovation on the level of physical storage, it does not affect logical schema or query semantics, only the cost analysis of each query plan.

Similar to having an index on either join input by join columns, a query planner may consider a merged index as pre-sorted inputs and favors a merge join plan.
A merged index further reduces the cost of a merge join plan by storing all matching records close together in one storage structure.
Combined with factors such as whether the merged index covers all columns in the query, join type (e.g., inner join or outer join), join selectivity, and availability of an equivalent materialized join view, a query planner may select a merged index-backed join plan for the current query.
Exact decision-making protocols and cost models differ among query planners, and we leave the exploration of such details to future work.

\subsection{Summary of techniques}\label{ssec:techniques-summary}

Merged indexes speed up index-based merge joins and answer join queries efficiently.
Here are the key benefits of merged indexes:

\begin{itemize}
    \item \textbf{Ideal query performance}: Merged indexes achieve query performance comparable to materialized join views. Matching records are stored close together, enabling efficient retrieval of join results.
    \item \textbf{Ideal maintenance efficiency}: Merged indexes provide maintenance efficiency similar to traditional single-table indexes. Updates to base tables are reflected in the merged index in a one-to-one manner.
    \item \textbf{Moderate space requirements}: A merged index is practically the same size as the sum of the equivalent traditional indexes on base tables. Unlike a materialized join, it does not require extra space in the database but reorganizes traditional indexes into a different structure. In addition, The merged index is in itself a \textbf{natural compression} of the outer join result.
    It retains one copy of each record and does not replicate a record if it matches multiple times.
    \item The merged index is effectively a \textbf{precomputed outer join}:
    It stores records from multiple tables and does not exclude records if they lack a match. It is \textbf{agnostic to the join type}, and one such storage structure supports all join types of the same input tables on the same columns or any prefix (in the case of composite keys).
\end{itemize}

\section{Experiments and Findings}\label{sec:exp}

This section evaluates the performance of merged indexes across a wide range of configurations.
\cref{ssec:techniques-summary} sets the agenda for the experiments:
\Cref{ssec:query} investigates the hypothesis of ideal query performance,
\Cref{ssec:maintenance} evaluates the hypothesis of ideal maintenance efficiency, and
\Cref{ssec:space} examines the hypothesis of moderate space requirements.

The results confirm these hypotheses and offer an unexpected finding: merged indexes can surpass traditional indexes in maintenance and materialized join views in query performance.
However, they also highlight trade-offs in scenarios like low join selectivity, advising caution in such cases.

\subsection{Experiment setup, workload, and variables}\label{ssec:setup}

The experiments focus on a binary join operator between the stock and orderline tables, as defined in the TPC-C benchmark~\cite{tpc2010benchmark}.
Three storage structures---merged indexes, traditional indexes, and materialized join views---are compared for join queries and update maintenance.
All storage structures use the same format, either b-trees or log-structured merge-forests. Materialized join views are stored as index-organized tables.
Both base tables are updated with the TPC-C new-order transaction, and each storage structure is maintained accordingly as part of the update transaction.
Two join queries are evaluated: a point lookup and a range scan of the join output.
The SQL definitions are as follows; Outer join experiments must slightly adjust this definition accordingly.

{\footnotesize
\begin{verbatim}
SELECT o.*, quantity, yearToDate, orderCount, data
            -- all stock columns except 10 text fields
FROM Orderline as o, Stock as s
WHERE o.itemID = s.itemID AND o.warehouseID = s.warehouseID
    -- for point lookup: AND warehouseID = ? AND itemID = ?
    -- for range scan: AND warehouseID = ?
\end{verbatim}
}

The experiments consider six independent variables (\cref{tab:variables}) and report results from 189 experiments across a wide range of scenarios.

\begin{small}
\begin{longtable}{ | r | p{0.6\linewidth} | l | }
    \caption{Independent variables of our experiments.}\label{tab:variables} \\
    \hline
    Variable Name & Description and/or Experiment Options & Default \\
    \hline
    \endhead
    \multicolumn{3}{|c|}{\textbf{System-dependent Variables}} \\*
    \hline
    \textbf{Buffer pool size} & Two options: 1 GB (\textit{beyond-memory} workload) or 16 GB (\textit{in-memory} workload) & 1 GB \\*
    \textbf{Storage format} & Two options: b-tree (LeanStore~\cite{leanstore18}) or log-structured merge-forest (RocksDB~\cite{rocksdb21}) & B-tree \\
    \hline
    \multicolumn{3}{|c|}{\textbf{Database-dependent Variable}} \\*
    \hline
    \textbf{Join selectivity} & Selectivity of \texttt{ORDERLINE SEMI JOIN STOCK}, abbreviated as \(SO\). $SO=100\%$ indicates referential integrity. At $SO=19\%$, storage structure sizes are roughly equal using default values for other parameters. & 100\% \\
    \hline
    \multicolumn{3}{|c|}{\textbf{Query-dependent Variable}} \\*
    \hline
    \textbf{Join type} & Two options: inner join or (full) outer join. & Inner join \\
    \hline
    \multicolumn{3}{|c|}{\textbf{Policy-dependent Variable}} \\*
    \hline
    \textbf{Included columns} & Specifies which columns from input tables are stored in the storage structures. Options: all columns (``all''), only covering columns of the two known join queries (``covering''), or primary key columns only (``keys''). & all \\
    \hline
\end{longtable}
\end{small}

The experiments run on a Google Cloud c2-standard-8 VM instance (8~vCPUs, 32~GB RAM, 200~GB SSD-backed storage) with Ubuntu 22.04.
Three worker threads simulate background read-only transactions on TPC-C tables, while one worker thread runs the join queries or update transaction.
Each experiment leads with an initial warm-up stage, which includes log-structured merge-forest compaction.

\subsection{Performance of queries: point lookups and range scans}\label{ssec:query}

\paragraph{Merged indexes outperform merge joins with traditional indexes.}

\begin{figure}[htb]
    \centering
    \begin{subfigure}{\linewidth}
        \centering
        \resizebox{\linewidth}{!}{\subimport{img}{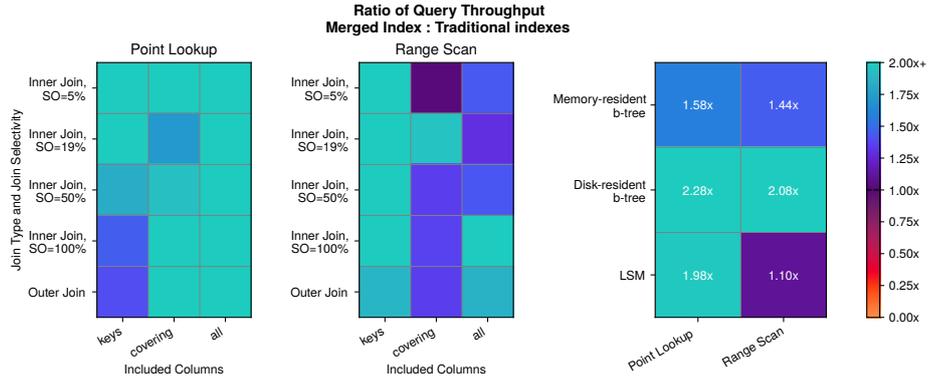}}
        \caption{\textbf{Merged index vs. traditional indexes}:
            Merged indexes achieve significant speedup.}\label{fig:query_speedup}
    \end{subfigure}

    \begin{subfigure}{\linewidth}
        \centering
        \resizebox{\linewidth}{!}{\subimport{img}{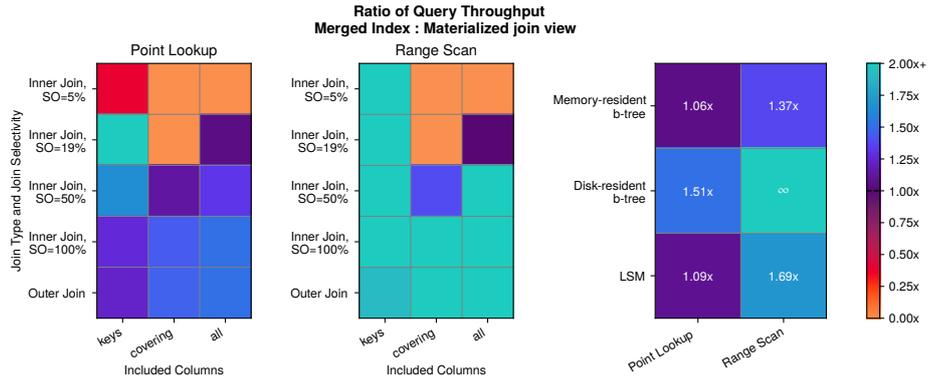}}
        \caption{\textbf{Merged index vs. materialized join view}:
            Merged indexes generally match materialized join views, with exceptions in red and orange cells.
    }\label{fig:query_match}
    \end{subfigure}
    \caption{\textbf{Query performance of merged index, traditional indexes, and materialized join views}, reported as query throughput ratios.}\label{fig:query_all}
\end{figure}

\Cref{fig:query_speedup} compares query performance of merged and traditional indexes across 36 cases.
Merged indexes match or surpass traditional indexes, achieving more than 2x query throughput in most cases.
This speedup is more pronounced in point lookups: with traditional indexes, queries must search two separate B-trees, whereas a merged index allows a single search that returns results for both tables. For range scans, on the other hand, the difference is smaller because the cost of two separate lookups in traditional indexes is amortized as the scan continues.

\paragraph{Merged indexes generally match materialized join views in query performance.}

\Cref{fig:query_match} shows that query performance of merged indexes match materialized join views in 29 of 36 cases.
In common settings with referential integrity enforced between joined tables (\(SO=100\%\)), merged indexes outperform materialized join views by up to 4.1x.
The reason for even higher query throughput lies in the compression effect of the merged index compared with materialized join views.
Smaller sizes reduce number of comparisons needed to traverse the b-tree for point lookups and data volume for range scans.

In cases of low join selectivity, merged indexes incur overhead from retaining non-matching tuples, explaining the red and orange cells.\footnote{
    In the orange cells that indicate a throughput ratio of nearly zero, the materialized inner join is small enough to fit in memory.
}
Retaining non-matching tuples allows the same merged index to handle outer joins effectively.
If such outer join queries are uncommon in the workload, a materialized join view may be preferable.
Two further alternatives are a merged index with only matching rows and a merged index that separates, in the manner of partitioned b-trees, matching and non-matching rows.

\paragraph{CPU and I/O analysis of join queries}

\begin{figure}[!htb]
    \centering
    \resizebox{\linewidth}{!}{\subimport{img}{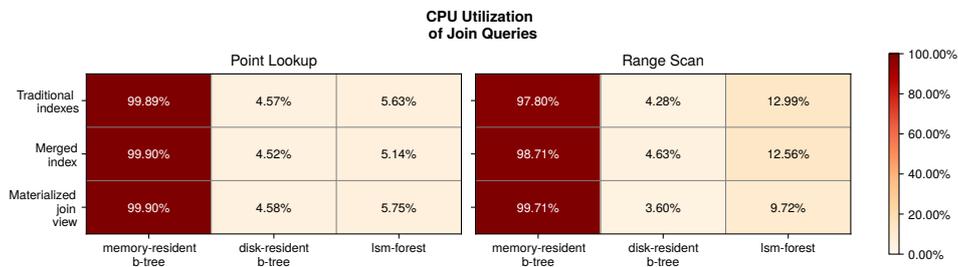}}
    \caption{\textbf{CPU utilization of join queries} for three approaches}\label{fig:cpu_utilization}
\end{figure}

The default buffer pool size of 1~GB creates an I/O-bound workload that saturates the external storage bandwidth.\footnote{
  Google Cloud's balanced persistent storage~\cite{gclouddisk} is backed by SSD but offers a slightly lower throughput than local SSDs but higher than HDDs.
}
For all external storage formats, CPU utilization of join queries remains below 15\%, as shown in \cref{fig:cpu_utilization}.
Though not shown in the figure, experiments evaluating update performance show similar CPU utilization metrics.

A merged index is better optimized for I/O-bound workloads than traditional options.
For point lookups, it halves b-tree searches compared to traditional indexes, while, compared to materialized join views, the compression effect of merged indexes reduces number of comparisons needed to traverse the b-tree.
This compression effect also reduces data volume of range scans needed for merged indexes compared to materialized join views.

Increasing the buffer pool size to 16~GB leaves the workload CPU-bound, with CPU utilization reaching 100\%.
While this reduces the performance advantage of the merged index, it still retains a significant edge over traditional options as shown in \cref{fig:query_speedup,fig:query_match,fig:maintenance}.

\subsection{Incremental maintenance and bulk updates}\label{ssec:maintenance}

\paragraph{Merged indexes match traditional indexes and outperform materialized join views in maintenance efficiency.}

\begin{figure}[htb]
    \centering
    \resizebox{\linewidth}{!}{\subimport{img}{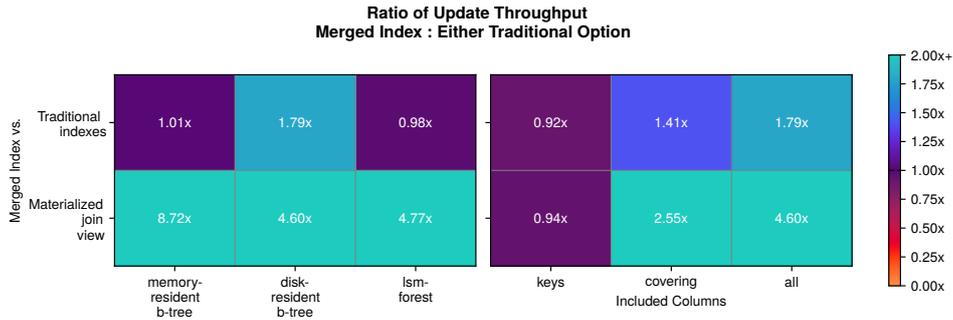}}
    \caption{
        \textbf{Maintenance efficiency of merged index, traditional indexes, and materialized join view}, reported as update throughput ratios.
        The merged index roughly matches traditional indexes and significantly outperforms materialized join views.}\label{fig:maintenance}
\end{figure}

\Cref{fig:maintenance} compares maintenance efficiency across storage structures.
The merged index achieves an update throughput up to 8\% lower but up to 79\% higher than traditional indexes, as shown in the upper row.
Its throughput surpasses that of materialized join views in nearly all cases, with throughput ratios reaching up to 8.72x, as shown in the lower row.

\paragraph{Impact of log-structured merge-forests on maintenance efficiency.}

Surprisingly, log-structured merge-forests do not always improve maintenance efficiency over b-trees.
As shown in \cref{tab:lsm_speedup}, log-structured merge-forests slow down maintenance for traditional and merged indexes by more than 25\% in the default setting.
As a matter of fact, update transactions are read-write transactions and quite heavy on reads---b-trees already spend over 95\% of disk accesses on reads.
The slower reads of log-structured merge-forests outweighs their faster write speeds in these cases.

\begin{table}[htb]
    \centering
    \caption{\textbf{Maintenance speedup or slowdown of log-structured merge-forests compared to b-trees.}
        TPut ratio: throughput ratio of log-structured merge-forests to b-trees.
        Read PCT: percentage of read accesses among all disk accesses.}\label{tab:lsm_speedup}
    \begin{tabular}{c || c || c c}
    \hline
    & \textbf{TPut ratio} & \textbf{Read PCT, b-trees} & \textbf{Read PCT, lsm-forests}\\
    \hline
    \textbf{Traditional indexes} & \textbf{1.34x} & 98.00\% & 99.96\%\\
    \textbf{Merged index} & \textbf{0.73x} & 96.60\% & 99.95\%\\
    \textbf{Materialized join view} & \textbf{0.71x} & 99.05\% & 99.95\%\\
    \hline
    \end{tabular}
\end{table}

\noindent However, log-structured merge-forests greatly reduce write time (read ratios consistently exceed 99.9\%) and may perform better in more write-heavy workloads.
They offer other benefits:
Experiments shown in \cref{fig:size_overview} show that their longer runs compress data effectively, reducing scan volume.
In addition, they provide efficient bulk loading and deletion, as discussed next.

\paragraph{Bulk loading records into merged indexes.}

\begin{figure}[!htb]
    
    \begin{minipage}{0.48\linewidth}
        \centering
        \resizebox{\linewidth}{!}{\subimport{img}{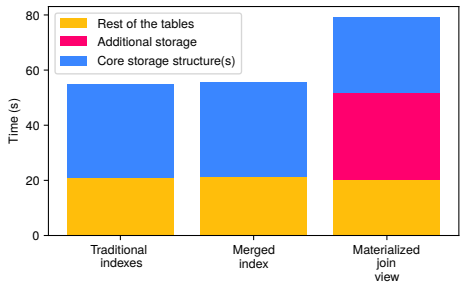}}
    \end{minipage}\hfill
    \begin{minipage}{0.48\linewidth}
    \caption{\textbf{Loading time of the full database in log-structured merge-forests}.
    Blue solid bars display the loading time of core storage structures, while the hatched bars show the loading time of other tables in the database.
        Records of two tables are loaded into a merged index without sorting first, simulating index creation.
        A merged index only loads slightly slower than equivalent single-table indexes.
    }\label{fig:lsm_load_time}
    \end{minipage}
\end{figure}

One drawback of merged indexes is less efficient index creation and dropping compared to traditional indexes.
While each single-table index can be dropped or added independently, a merged index must perform bulk deletion or insertion operations.
Log-structured merge-forests greatly mitigate their cost by deferring sorting or consolidating tombstone records until the next compaction run.
Experiments in \cref{fig:lsm_load_time} load records from to tables into the merged index (as a log-structured merge-forest) without sorting first, simulating this scenario.
Loading merged index is only slightly slower than loading two single-table indexes.
This addresses a potential drawback of merged indexes and makes them appealing for workloads with frequent index creation and dropping as well.

\subsection{Space requirements}\label{ssec:space}

\paragraph{The merged index adds minimal space overhead compared to traditional indexes.}

\begin{figure}[htb]
    \centering
    \resizebox{\linewidth}{!}{\subimport{img}{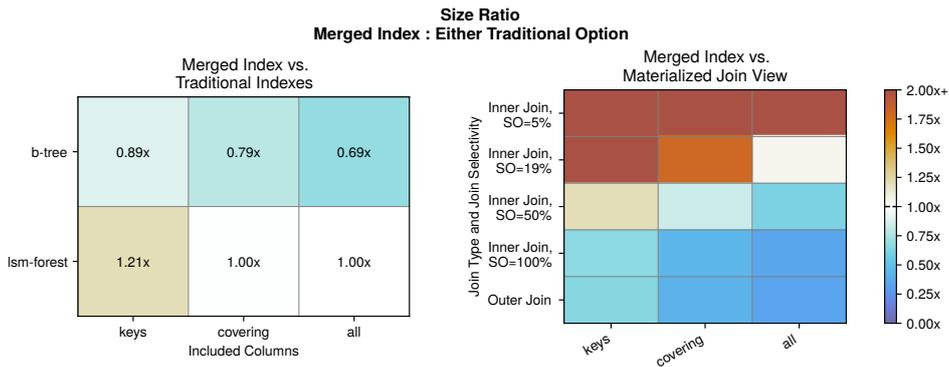}}
    \caption{
    \textbf{Space required by merged indexes, traditional indexes, and materialized join views.}
    Blue cells: merged index requires less space than traditional indexes. Off-white cells: similar space requirements. Red cells: merged index requires more space.}\label{fig:space_requirements}
\end{figure}

The left panel of \cref{fig:space_requirements} shows that merged indexes require between 69\% and 121\% of the space used by traditional indexes across six settings.
Log-structured merge-forests increase the relative space requirements of merged indexes.
As one possible reason, traditional indexes store only one record type, enabling more prefix compression in longer runs of the log-structured merge-forest.

\paragraph{Merged indexes naturally compress materialized join views in many settings.}

The right panel of \cref{fig:space_requirements} compares the merged index with materialized join views.
In most settings, including common configurations where \(SO=100\%\) and referential integrity holds, merged indexes require less space by storing only one copy of each record, regardless of the number of matches in the join.

However, the compression effect diminishes with fewer included columns, as duplicating smaller records in materialized join views is less costly.
Also, low join selectivity increases the space overhead incurred by retaining non-matching records.
Red and orange cells indicate that the space overhead outweighs the compression effect.
As discussed in \cref{ssec:query}, the merged index may not be the best choice in those scenarios if the database does not also process outer join queries between the two tables.

\paragraph{Space requirements for the full database.}

\begin{figure}[htb]
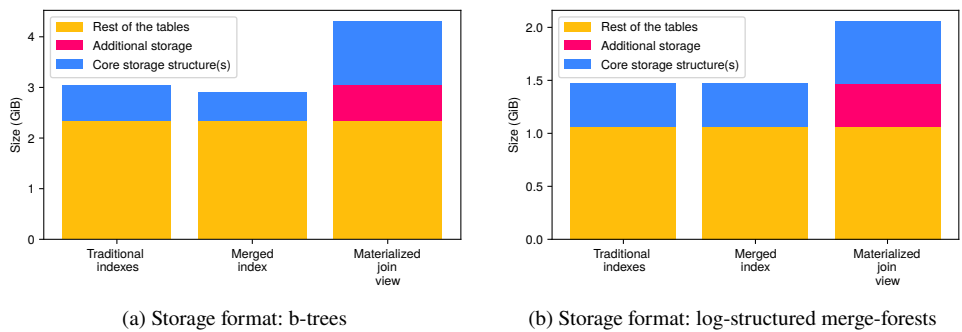

    \begin{subfigure}{0.48\linewidth}
        \centering
        \resizebox{\linewidth}{!}{\subimport{img}{size_b-tree.pgf}}
        \caption{Storage format: b-trees}\label{fig:size_b-tree}
    \end{subfigure}\hfill
    \begin{subfigure}{0.48\linewidth}
        \centering
        \resizebox{\linewidth}{!}{\subimport{img}{size_lsm-forest.pgf}}
        \caption{Storage format: log-structured merge-forests}\label{fig:size_lsm-forest}
    \end{subfigure}
    \caption{\textbf{Space required for the full database} across three approaches, with included columns set to ``covering.''
    Solid blue bars: storage structure size. Dashed yellow bars: base table size. Dashed pink bars: additional space for materialized join views.}\label{fig:size_overview}
\end{figure}

\Cref{fig:size_overview} illustrates the space requirements for the full database with covering columns included.
Materialized view is inefficient in space usage: First, it duplicates records that are matched multiple times in the join, inflating the size of the materialized join view.
Second, materialized join views require additional space to store secondary indexes of orderline and stock tables to speed up lookup of matches.
These indexes are necessary for the efficient maintenance of materialized join views.
This additional storage is effectively equal in size to traditional indexes or the merged index. 
Materialized join views, in itself space consuming already, are a net addition to the space required by the other two approaches.

\section{Summary and Conclusions}\label{sec:sum}

The motivation for this study is the need for efficient non-blocking join processing as well as efficient maintenance of required intermediate storage structures.
Previous studies face a trade-off between efficient query execution and efficient maintenance.
On one hand, materialized join views set the gold standard for efficient query processing but incur significant maintenance overhead;
on the other hand, traditional indexes set the gold standard for efficient maintenance but suffer from less efficient query execution.
Additionally, in contrast to most approaches to materialized join views, traditional indexes provide query coverage for inner join, all outer joins, and all semi joins.
This study introduces the merged index for this purpose, a storage structure that combines the advantages of both traditional single-table indexes and materialized two-table join views.

\begin{figure}[htb]
    \resizebox{\linewidth}{!}{\subimport{img}{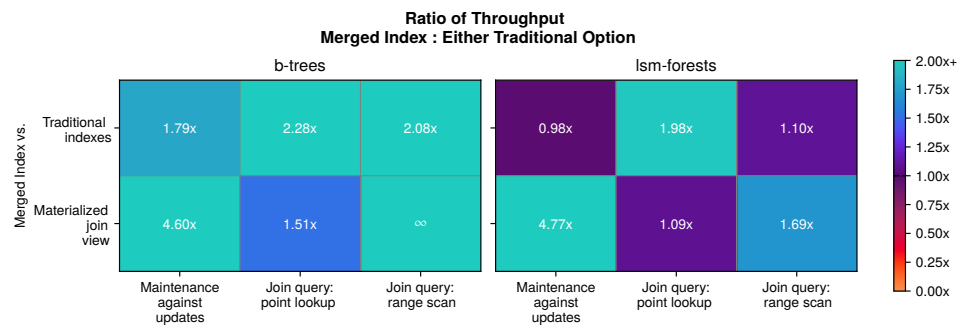}}
    \caption{\textbf{Relative transaction throughput of merged index and traditional alternatives in a typical scenario: an inner join with referential integrity.}
        In point lookups, range scans, and updates, merged indexes process up to 4.8x as many transactions as either traditional alternative.} \label{fig:perf}
\end{figure}

Experiments in \cref{sec:exp} compare the merged index with the two traditional options across a wide range of configurations with altogether 189 individual experiments.
\Cref{fig:perf} summarizes the performance for the most common scenario: an inner join with referential integrity, while \cref{fig:size_overview} compares space requirements.
These figures together support the following summary findings about merged indexes:

\begin{itemize}
    \item \textbf{Ideal maintenance efficiency}: As shown in the left column of \cref{fig:perf}, merged indexes are updated as efficiently as equivalent indexes on the join inputs and significantly more efficiently than materialized join views.
    \item \textbf{Ideal query efficiency}: The center and right columns of \cref{fig:perf} show that merged indexes can produce join results with efficiency comparable to or better than materialized views, and far more efficiently than indexes on join inputs.
    \item \textbf{Moderate space requirements}: \cref{fig:size_overview} shows that a merged index requires space nearly equal to that of equivalent individual indexes. A merged index effectively is a natural compression of the equivalent join view.
\end{itemize}

As a merged index of two database tables encompasses their full outer join, it is possible to extract either table but not as efficient as scanning a single-table storage structure.
As a merged index interleave both key values of the inner join and non-matching join key values, extracting the inner join from a full outer join again may be less efficient than scanning a storage structure holding the inner join only.

In conclusion, physical data independence, a hallmark of the relational approach to database management, permits  using any storage structure that represents and maintains logical tables and their contents, including the merged index.
A storage structure with multiple record types does not violate or undermine the principles of the relational approach. Accordingly, several important data storage systems prove its industrial viability~\cite{ oracle-table-cluster, Google-Spanner}; query processing using these storage systems is also compatible with industrial-strength query optimization.

\printbibliography
\end{document}